\documentclass[12pt]{article}

\usepackage{amsmath}
\usepackage{amsfonts}
\usepackage{amscd}
\usepackage{graphicx}

\begin{document}

\title{Hierarchical model of the actomyosin molecular motor based on ultrametric diffusion with drift}

\author{Andrei Khrennikov\\
International Center for Mathematical
Modelling \\in Physics and Cognitive Sciences \\
Linnaeus University,  Sweden\\
Sergei Kozyrev\\
Steklov  Mathematical Institute\\ Moscow, Russian Federation\\
 Alf M\aa nsson\\
School of Natural Sciences\\
Linnaeus University,  Sweden}

\maketitle

\begin{abstract}
We discuss the approach to investigation of molecular machines using systems of integro--differential ultrametric ($p$-adic) reaction--diffusion equations with drift. This approach combines the features of continuous and discrete dynamic models. We apply this model to investigation of actomyosin molecular motor.

The introduced system of equations is solved analytically using $p$-adic wavelet theory. We find explicit stationary solutions
and behavior in the relaxation regime.
\end{abstract}

\section{Introduction}

In the present paper we develop the approach to description of molecular machines (namely actomyosin motor) from the point of view of models of hierarchical dynamics in complex systems.

Models of hierarchical dynamics in application to protein dynamics were discussed starting from 1980-ies, cf. \cite{Frauenfelder4}, \cite{Hatom} \cite{Frauenfelder5}, \cite{FSW}. In these models the protein dynamics was described by transitions in a hierarchy of states. This approach was motivated by the theory of spin glasses \cite{MPV}, \cite{FischerHertz}, \cite{BinderYoung} and by the methods of nonequilibrium dynamics \cite{Bak}. In particular in \cite{Frauenfelder4} the two types of motions in proteins were discussed --- fast equilibrium fluctuations (EF) and slower functionally important motions (FIM).
The dynamics of proteins was described by sequences of proteinquakes --- avalanche processes of conformational rearrangements, in analogy to the processes used for description of earthquakes \cite{Bak}.

Application of the so called fractal (or crumpled) polymer globules to molecular machines was discussed in \cite{AvetisovNechaev}, \cite{crumpled}. It was mentioned that due to topological constraints on the space of conformations of polymers in crumpled globules the spectra of conformational transitions are discrete. This implies the existence of the energy gap between the fast and slow degrees of freedom. This allows to represent the dynamics of the crumpled globule as a comparably fast relaxation to an attractor of low dimension. Dynamics of this attractor will describe the molecular machine.

A characteristic feature of a machine is low dimension of the space of states. Space of states of a protein has a very high dimension. Therefore the description of the mechanism of emergence of effective low dimensional dynamics is important for the investigation of molecular machines.

Discrete spectrum of the above mentioned dynamics of the fractal polymer globule is described by the block--hierarchical Parisi matrix. Matrices of this kind arise in theory of complex systems (spin glasses) and they are used for description of spaces of states of glasses \cite{MPV}. In \cite{ABK}, \cite{PaSu} it was shown that the Parisi matrix possesses a natural $p$-adic (ultrametric) parametrization. In application to the models of hierarchical dynamics of macromolecules this implies that this dynamics can be modelled by $p$-adic diffusion. In \cite{ReplicaI,ReplicaII,ReplicaIII} a generalization of the $p$-adic approach to spin glasses onto more general ultrametric spaces was developed.

Let us note that models of $p$-adic (ultrametric) diffusion were initially developed from purely mathematical motivations \cite{VVZ}, \cite{AlbKarw}.

From the point of view of dynamics on landscapes of energy the discussed hierarchical dynamics can be considered as a sequence of Arrhenius transitions in the hierarchy of local energy minima and transition states. This kind of dynamics is equivalent to some model of ultrametric diffusion \cite{Landscape}. The energy landscape for short peptides was investigated in \cite{BeckerKarplus} and the corresponding hierarchy of states (disconnectivity graph of the energy landscape) was constructed.

Let us stress that an ultrametric diffusion differs from a real anomalous diffusion by a discrete spectrum of the generator. This property allows the existence of the energy gap which separates the degrees of freedom of the molecular machine from faster conformational rearrangements.

In the present paper we use for the description of the molecular machine (namely the actomyosin motor) a system of equations of ultrametric reaction--diffusion. It is important to note that we use the equations of ultrametric diffusion with drift.

The simplest model of ultrametric diffusion has the form \cite{ABK,ABKO}
$$
\frac{\partial}{\partial t} f(x,t)+D^{\alpha}_{x}f(x,t)=0,
$$
where $t\in\mathbb{R}$ is time, $x\in\mathbb{Q}_p$ in application to dynamics of macromolecules is the variable which describes the space of conformations of the molecule by the hierarchy of local minima and transition states of the energy landscape. The operator $D^{\alpha}_{x}$ (the generator of ultrametric diffusion) is the Vladimirov operator of $p$-adic fractional differentiation which desribes transitions in the hierarchy of states. The index $\alpha>0$ is proportional to the inverse temperature $\beta$.

The above equation describes hierarchical dynamics for the case when all local minima for the energy landscape have equal energy. This approximation is sufficiently realistic for spin glasses but for molecular machines one has to take into account transitions between the local minima with different energy. The corresponding modification of the diffusion generator (generator of ultrametric diffusion with drift) has the form $D^{\alpha}_{x}\cdot e^{\beta U(x)}$ of  product of the diffusion generator and  operator of multiplication by the function $e^{\beta U(x)}$ where $U(x)$ describes the energies of the corresponding local minima and $\beta$ is the inverse temperature.

Molecular motors and their applications are hot topics in molecular biology and biophysics, see, e.g., \cite{M2}, \cite{M3}.
Recently actomyosin molecular motors have attracted  a lot of attention in applied research \cite{M3a}, \cite{M3b}, \cite{M4}, \cite{M5}. Models based on novel approaches can
have impacts to better understanding of functioning of such motors. One of such models is presented in this paper.

In order to relate observed cellular events due to protein machines to the dynamic properties of the machines in the ensemble, one may use statistical models that take into account the key features of the molecular energy landscapes. In spite of more recent developments, e.g., the use of molecular dynamics simulations \cite{Li}, \cite{Yu} and various coarse grain structural simulation approaches \cite{Craig}, there is still considerable interest in the statistical modelling based on suitably defined energy landscapes and key conformational states without detailed structural information. These types of models thus focus on key physical properties without blurring the big picture by details, they can readily be adapted to time scales relevant to molecular machine function in the cell (not always the case with, e.g., molecular dynamics simulations) and the modelling is computationally cheap.

However, whereas being considerably simpler than molecular dynamics simulations, most statistical models of muscle contraction, e.g., Duke \cite{Duke},  Smith et al. \cite{Smith1}, \cite{Smith2},  Albet-Torres et al 2009; M\aa nsson \cite{Albet-Torres}, \cite{{Mansson}},
Walcott and  Warshaw \cite{Walcott}, \cite{Walcott1} are not readily solved analytically but require Monte-Carlo approaches or use various numerical methods.

Finally, we remark that one may point out that (for standard real models of molecular motors) the solutions of reaction diffusion equations
are typically not very difficult to find numerically, e.g.,  Lan and Sun \cite{Landscape1}, Oster and  Wang \cite{Landscape2}.  Therefore one may question
the need of development of novel mathematical models (e.g., based on $p$-adics) which provide the possibility to obtain
analytical solutions. However, the reader will see that by getting analytical solutions we are able to get finer results on the mathematical structure of the model. For example, by moving to the $p$-adic domain and deriving analytical solutions we describe
in details the process of relaxation to the stationary solution. We point also to another important (and advantageous) feature of the $p$-adic model. Here the model determining functions
$k_i(x)$ (reaction rates) and $U(x)$ (potentials) can be chosen very simply
(as scalings of characteristic functions of $p$-adic balls). Already this choice implies very natural (from the biological viewpoint) dynamics.

The exposition of this paper is as follows.

In Section 2 we discuss basic constructions of ultrametric analysis.

In Section 3 we discuss ultrametric models of landscape dynamics and application to dynamics of macromolecules.

In Section 4 we construct the model of the molecular motor based on the system of ultrametric diffusion equations with drift.

In Section 5 we find the exact solution for the stationary state of the reaction--diffusion model constructed in Section 4.

In Section 6 we describe analytically the relaxation to the stationary state found in Section 5.

\section{Ultrametric spaces and $p$-adic numbers}

In the present section we discuss in short the definition and some properties of ultrametric spaces, in particular fields of $p$-adic numbers.
For introduction to $p$-adic analysis see \cite{VVZ}, \cite{Gouvea}.

A metric $d(\cdot,\cdot)$ satisfies
the strong triangle inequality (and called an ultrametric) if
$$
d(x,y)\le {\rm max}(d(x,z),d(y,z)),\qquad \forall x,y,z.
$$

Ultrametric space possesses the following properties:

1) All triangles are isosceles;

2) two balls either do not intersect or one of the balls contains the other;

3) balls are hierarchically nested
(i.e. we have a partially ordered hierarchical tree of balls).

\medskip

An important example of an ultrametric space is given by the filed of $p$-adic numbers, defined as follows.
$p$-Adic norm $|\cdot|_p$ of a rational $x$ is given by the following expression: for a prime $p$
$$
x=p^{\gamma}{m\over n},\quad |x|_p=p^{-\gamma},\qquad |0|_p=0,
$$
where integers $m$ and $n$ are not divisible by $p$.

The field $\mathbb{Q}_p$ of $p$-adic numbers is a
completion of $\mathbb{Q}$ with respect to $p$-adic norm.

$p$-Adic numbers are in the one to one correspondence with series
$$
x=\sum_{i=\gamma}^{\infty}x_ip^i,\quad x_i=0,\dots,p-1
$$
where $\gamma$ is an arbitrary integer.

The ring $\mathbb{Z}_p$ of $p$-adic integers is
the unit ball: $|x|_p\le 1$. This ball contains
all the above series over the degrees of $p$ with $\gamma\ge 0$.

We denote $\Omega(\cdot)$ the characteristic function of $[0,1]$, thus $\Omega(|\cdot|_p)$ is the characteristic function of $p$-adic unit ball $\mathbb{Z}_p$.

The $p$-adic wavelet $\psi_J$ associated to a ball $J$ in $\mathbb{Q}_p$ is a mean zero complex valued function with the support in $J$ such that $\psi_J$ is constant on maximal subballs in $J$ (i.e. $\psi_J$ is a mean zero linear combination of characteristic functions of maximal subballs in $J$). There exist $p-1$ linearly independent wavelets associated to each ball in $\mathbb{Q}_p$.

The Vladimirov operator of $p$-adic fractional differentiation has the form
\begin{equation}\label{PDO}
D^{\alpha}f(x)=\Gamma_p^{-1}(-\alpha)\int_{\mathbb{Q}_p}{f(x)-f(y)\over
|x-y|_p^{1+\alpha}}d\mu(y)
\end{equation}
where $\mu$ is the Haar measure in $\mathbb{Q}_p$.

$p$-Adic wavelets are eigenvectors of pseudodifferential operators \cite{wavelets}, in particular
$$
D^{\alpha}\psi_J=p^{\alpha(1-\gamma)}\psi_J
$$
where $p^{\gamma}$ is the diameter of the ball $J$.

\section{Ultrametric methods of dynamics on complex energy landscapes}

Dynamics on complex energy landscapes was discussed in relation to spin glasses and dynamics of proteins \cite{FSW}. One of the approaches is related to the approximation of the dynamics by the system of kinetic equations which describe transitions between local minima of energy. System of kinetic equations of this kind has the form
\begin{equation}\label{kinetics}
{d\over dt}f_{a}(t)=-\sum_{b}Q_{ab}(f_{a}(t)-f_{b}(t)),
\end{equation}
where $t$ is time, $f_{a}(t)$ is the population of the $a$-th local minimum, $Q_{ab}$ are transition rates.

The approach of hierarchical (or basin to basin) kinetics for dynamics on complex energy landscapes is based on the following observation: for
three local minima of the energy landscape in general there exist only two transition states (saddle points of the energy landscape).

Therefore the dynamics of transitions between the local minima will be defined by a hierarchy of transition states and in some approximation will be described by some model of hierarchical (or ultrametric) kinetics. Groups of local minima separated by energy barriers are called basins and the corresponding systems of kinetic equations are called models of interbasin (or basin to basin) kinetics.
In this hierarchical approximation the matrix of transition rates $(Q_{ab})$ in (\ref{kinetics}) will be given by the Parisi hierarchical block matrix discussed in the theory of spin glasses \cite{MPV}
$$
(Q_{ab})=
\begin{array}{c}
\begin{array}{|cc|cc|cccc|}
\hline
{0}&q_{1}&q_{2}&q_{2} &q_{3}&q_{3}&q_{3}&q_{3}\\
q_{1}&{0}&q_{2}&q_{2}&q_{3}&q_{3}&q_{3}&q_{3}\\
\cline{0-3}
q_{2}&q_{2}&{0}&q_{1}&q_{3}&q_{3}&q_{3}&q_{3}\\
q_{2}&q_{2}&q_{1}&{0}&q_{3}&q_{3}&q_{3}&q_{3}\\
\hline
\end{array}
\\
\begin{array}{|cccc|cc|cc|}
q_{3}&q_{3}&q_{3}&q_{3}&{0}&q_{1}&q_{2}&q_{2}\\
q_{3}&q_{3}&q_{3}&q_{3}&q_{1}&{0}&q_{2}&q_{2}\\
\cline{5-8}
q_{3}&q_{3}&q_{3}&q_{3}&q_{2}&q_{2}&{0}&q_{1}\\
q_{3}&q_{3}&q_{3}&q_{3}&q_{2}&q_{2}&q_{1}&{0}\\
\hline
\end{array}
\end{array},
$$
where $q_i=e^{-\beta E_i}$ are positive parameters (the Arrhenius transition rates, $E_i$ are the corresponding energy barriers and $\beta$ is the inverse temperature) and in the above formula an example of $8\times 8$ Parisi matrix is shown. This hierarchical dynamics approach was discussed in many papers, in particular in \cite{OgielskyStein}, \cite{HoffmannSibani}, \cite{BlumenKlafter}. We are interested in the regime of infinitely complex rugged landscape of energy which in this approach is described by infinite Parisi matrices. The corresponding systems of kinetic equations give rise to models of ultrametric diffusion \cite{ABK}.

$p$-Adic parametrization of the Parisi matrix was proposed in  \cite{ABK}, \cite{PaSu}. In \cite{ReplicaI}, \cite{ReplicaII}, \cite{ReplicaIII} more general ultrametric replica solutions were proposed.

Ideas of the theory of spin glasses were applied by Frauenfelder and coathors to description of Myoglobin--CO rebinding   \cite{Frauenfelder4}, \cite{Hatom}, \cite{Frauenfelder5}. In  paper \cite{BeckerKarplus} by Becker and Karplus hierarchies of local minima (''disconnectivity graph'' of basins of energy landscape) for short peptides were constructed using numerical simulation.


Let us review several examples of application of $p$-adic diffusion to dynamics of macromolecules.

\medskip

\noindent{\bf Example.  }\quad (See also \cite{ABK}).
Let us consider the equation of $p$-adic diffusion
\begin{equation}\label{protein}
{\partial\over\partial t}f(x,t)+D_x^{\alpha}f(x,t)=0
\end{equation}
generated by the Vladimirov fractional operator.
The parameter $\alpha$ is proportional to the inverse temperature: $\alpha=\lambda\beta $.

This equation describes the protein dynamics (conformational diffusion).
The $p$-adic coordinate $x$ parameterizes the set of conformations of a protein. In the language of \cite{BeckerKarplus} the above equation describes the diffusion on the border of the disconnectivity graph of landscape of energy.

The diffusion generator $D^{\alpha}_x$ generates jump process, i.e. trajectories of the corresponding diffusion are discontinuous. In the experiments we observe the dynamics of proteins (in particular, molecular motors) at time scales which are large in comparison to microscopic dynamics. At large time scales protein dynamics takes the form of ''proteinquakes'' \cite{Frauenfelder4}, or jumps between local minima of the energy landscape. In application to molecular motors this will give discrete models of molecular motors, discussed for example in  \cite{M2}.

\medskip

\noindent{\bf Example. Ultrametric reaction--diffusion equation: binding Mb--CO.}\quad (See also \cite{ABKO}).
Binding Mb--CO is possible when the myoglobin molecule is in a conformation
with open path to iron inside the myoglobin globule.
We describe this set of conformations by a unit ball in $\mathbb{Q}_p$.

The model of Mb--CO binding is described by the
equation of $p$-adic diffusion with a sink
(ultrametric reaction--diffusion equation)
$$
\left[{\partial\over
\partial t}+D_x^{\lambda\beta}+\Omega(|x|_p)\right]f(x,t)=0.
$$
Here $\Omega(|x|_p)$ --- characteristic function of the unit ball $|x|_p\le 1$,
$f(x,t)$ --- function of distribution over conformations of myoglobin molecules not bound to CO.

This model reproduces the results of the experiments by Frauenfelder on Mb--CO binding dynamics \cite{Frauenfelder4,Hatom,Frauenfelder5}.

\medskip

$p$-Adic model of a molecular machine described by a system of equations of bounded $p$-adic reaction--diffusion (without drift) was considered in \cite{AvBikZub}.
Some other results in application of ultrametric methods to protein
dynamics were obtained: a) different relaxation dependencies were described in \cite{A1}; b)
first passage time for ultrametric random walks were described
 in \cite{A2}; c) spectral diffusion by ultrametric approach was studied in \cite{A3}, d)
different applications of random hierarchic matrices were discussed in \cite{A4},  in
particular, applications to packing of DNA.

\section{Actomyosin molecular motor}

Let us discuss the actomyosin molecular motor, see for example \cite{M2}. In the present section we build a model of this molecular motor using the discussed in the previous section approach of ultrametric modeling of dynamics on complex landscapes of energy. Using ultrametric methods we take into account the complexity of the energy landscape and the anomalous character of the conformational diffusion of the protein molecules.

Actomyosin molecular motor performs motility in cells.
The scheme of reactions and conformational rearrangements of actomyosin molecular motor has the form described below. The working cycle of the motor includes two major conformational rearrangements of myosin, described by horizontal arrows at the scheme below.
$$
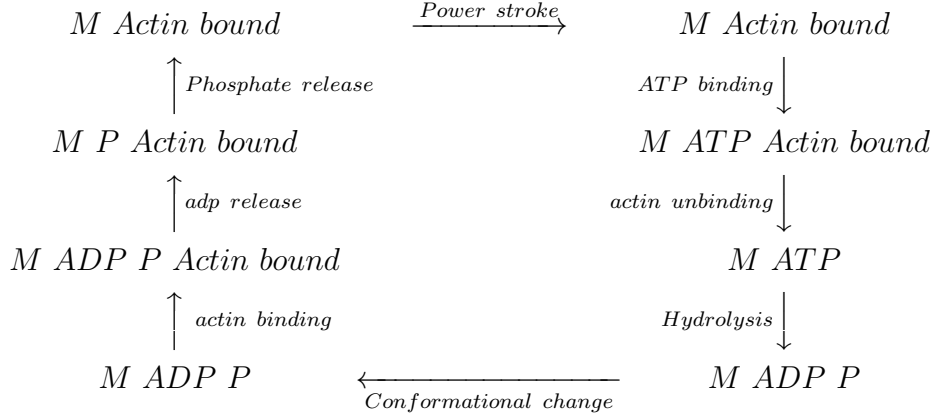
\begin{CD}
M~Actin~bound @> Power~stroke >>  M~Actin~bound\\
@AA Phosphate~release A @V ATP~binding VV\\
M~P~Actin~bound & & M~ATP~Actin~bound\\
@AA adp~release A @V actin~unbinding VV\\
M~ADP~P~Actin~bound & & M~ATP\\
@AA ~actin~binding A @V Hydrolysis VV\\
M~ADP~P @<< Conformational~change < M~ADP~P
\end{CD}
$$

In principle the description of the above sequence of transformations should include the conformational transformations in the process of binding and unbinding of actin to myosin. In order to simplify the model we ignore this (comparably fast and small) conformational rearrangement of myosin. With this simplification the above scheme will be described by the system of the two ultrametric reaction--diffusion equations with drift where the diffusion terms describe conformational transformations corresponding to the horizontal arrows in the scheme above
\begin{equation}\label{p-adic1}
{\partial\over \partial t} f_{1}(x,t)= -\left[k_{1}(x)+D_{x}^{\lambda\beta}e^{\beta U_{1}(x)}\right] f_1(x,t)+k_{2}(x) f_{2}(x,t);
\end{equation}
\begin{equation}\label{p-adic2}
{\partial\over \partial t} f_{2}(x,t)= k_{1}(x) f_1(x,t)-\left[k_{2}(x)+D_{x}^{\lambda\beta}e^{\beta U_{2}(x)}\right] f_{2}(x,t).
\end{equation}
The equations above imply the conservation law (conservation of number of myosin molecules)
\begin{equation}\label{p-adic3}
\int_{\mathbb{Q}_p}(f_{1}(x,t)+f_{2}(x,t))d\mu(x)={\rm const}.
\end{equation}

Here $f_{1}(x,t)$ (correspondingly $f_{2}(x,t)$) is the distribution over conformations $x$ of myosin bound to actin (correspondingly not bound to actin).
The $k_1(x)$ describes the conformationally dependent reaction rate of the following sequence of transformations of myosin: ATP binding, actin unbinding, ATP hydrolysis. The $k_2(x)$ is the reaction rate of actin binding, ADP and phosphate release. The $U_1(x)$ is the energy landscape of myosin bound to actin. The $U_2(x)$ is the energy landscape of a complex of myosin, ADP and phosphate where myosin is not bound to actin.

The operator $D_{x}^{\lambda\beta}e^{\beta U(x)}$ of ultrametric diffusion with drift has the form of the product of the ultrametric diffusion operator $D_{x}^{\lambda\beta}$ and the operator of multiplication by the function $e^{\beta U(x)}$ \cite{Landscape}.

In order to fix the above system of equations we make the following assumptions.

1) All reaction rates in the above system are proportional (with positive coefficients) to characteristic functions of some balls.

2) All potentials for the system above are proportional to characteristic functions of some balls with negative coefficients of proportionality (i.e. describe potential wells).

The corresponding reaction rates will be equal to:
$$
k_1(x)=k_1\Omega(|x|_p),\qquad k_{2}(x)=k_{2}\Omega(|x-a|_p),\quad |a|_p>1,
$$
where $k_1$, $k_2$ are positive.

We choose the potentials to be proportional to the same characteristic functions of balls as the corresponding reaction rates, i.e. to choose
$$
U_1(x)=U_1\Omega(|x|_p), \qquad U_{2}(x)=U_{2}\Omega(|x-a|_p),
$$
where $U_{1}$, $U_{2}$ are negative.

This choice corresponds to the following picture: the drift (conformational dynamics) goes in the direction of the potential well. The reaction takes place in this potential well. After the reaction the myosin molecule arrives to the different potential surface (which corresponds to the different complex of myosin with actin) and performs the second part of the cycle.

\section{Stationary solution for the molecular motor}

Let us investigate the stationary solution for the reaction--diffusion equations  (\ref{p-adic1}), (\ref{p-adic2}) for the molecular motor. We get
$$
D_{x}^{\lambda\beta}\left[e^{\beta U_{1}(x)}f_1(x)+e^{\beta U_{2}(x)}f_2(x)\right]=0,
$$
equivalently
\begin{equation}\label{sumGibbs}
e^{\beta U_{1}(x)}f_1(x)+e^{\beta U_{2}(x)}f_2(x)={\rm const}.
\end{equation}

Another equation for the stationary solution
\begin{equation}\label{waveletRHS}
D_{x}^{\lambda\beta}e^{\beta U_{1}(x)} f_1(x)=k_{2}(x) f_2(x) - k_{1}(x)f_1(x)
\end{equation}
implies that $k_{2}(x) f_2(x) - k_{1}(x)f_1(x)$ is a mean zero function, i.e.
$$
\int f_1(x)k_1(x)d\mu(x)=\int f_2(x)k_2(x)d\mu(x).
$$
The above condition has the physical meaning of the coincidence of the flows for the two stages of the work of the molecular motor.

\medskip

Let us choose the reaction rates in (\ref{p-adic1}), (\ref{p-adic2}) proportional to the characteristic functions of balls:
$$
k_1(x)=k_1\Omega(|x|_p),\qquad k_{2}(x)=k_{2}\Omega(|x-a|_p),\quad |a|_p>1,
$$
where $k_1$, $k_2$ are positive. Thus the reactions run in the two nonintersecting balls.

We construct the potentials $U_1(x)$, $U_2(x)$ as follows. These potentials correspond to potential wells proportional to the same characteristic functions of balls as the corresponding reaction rates $f_1(x)$, $f_2(x)$, and the potentials outside some ball are sufficiently large (for example, see below, are constant and equal to $U_{\infty}$).  We choose
$$
U_1(x)=U_1\Omega(|x|_p)+ U_{\infty}(1-\Omega(|p^{\gamma }x|_p)),
$$
$$
U_{2}(x)=U_{2}\Omega(|x-a|_p)+U_{\infty}(1-\Omega(|p^{\gamma }x|_p)),
$$
where $U_{1}$, $U_{2}$ are negative, $p^{\gamma}\ge |a|_p>1$ and $U_{\infty}>0$.

Let us take into account the above choice of the potential and the reaction rates.
For simplicity we also take $p^{\gamma}=|a|_p=p$, i.e. the reactions run in maximal subballs of a ball of the diameter $p$.
Then, up to multiplication by a constant, the RHS (right hand side) of (\ref{waveletRHS}) takes the form
\begin{equation}\label{waveletRHS_1}
k_{2}(x) f_2(x) - k_{1}(x)f_1(x)=\Omega(|x-a|_p)-\Omega(|x|_p).
\end{equation}
The above choice of the normalization constant corresponds to the unit flow.

The sketch of the proof is as follows. The function $k_{2}(x) f_2(x) - k_{1}(x)f_1(x)$ as a mean zero function possesses an expansion over wavelets. The wavelet with the largest support in this expansion has the form $\Omega(|x-a|_p)-\Omega(|x|_p)$. The expansion also may contain wavelets with smaller supports inside the balls with characteristic functions $\Omega(|x-a|_p)$, $\Omega(|x|_p)$. Substituting this expansion in (\ref{sumGibbs}), (\ref{waveletRHS}) and taking into account that for the operators of multiplication by functions $U_{1}(x)$, $U_{2}(x)$ the wavelets with sufficiently small supports are eigenfunctions, we get (\ref{waveletRHS_1}).

The function (\ref{waveletRHS_1}) is an eigenvector of the diffusion operator $D_{x}^{\lambda\beta}$ with the eigenvalue 1. Equation (\ref{waveletRHS}) with the normalization (\ref{waveletRHS_1}) implies the expression for the stationary state
\begin{equation}\label{f_1_stationary}
f_1(x)=e^{-\beta U_{1}(x)} \left(\Omega(|x-a|_p)-\Omega(|x|_p)+1+e^{\beta U_{1}}k_1^{-1}\right),
\end{equation}
\begin{equation}\label{f_2_stationary}
f_2(x)=e^{-\beta U_{2}(x)}\left(\Omega(|x|_p)-\Omega(|x-a|_p)+1+e^{\beta U_{2}}k_2^{-1}\right).
\end{equation}
In the limit $U_{\infty}\to +\infty$ we get the solution for $f_1$, $f_2$ localized in the ball with the diameter $p^{\gamma}$ with the center in zero.

\section{Relaxation to the stationary state}

In the present section we discuss the relaxation of the molecular motor described by the system (\ref{p-adic1}), (\ref{p-adic2}) of reaction--diffusion equations to the stationary state (\ref{f_1_stationary}), (\ref{f_2_stationary}). We consider the limit $U_{\infty}\to +\infty$ i.e. we put the system in the ball $|x|_p\le p$, moreover, we investigate the dynamics of the system in the space of locally constant functions with the support in this ball. Let us consider the dynamics in some subspaces of the mentioned space.

\medskip

Let us consider the dynamics (\ref{p-adic1}), (\ref{p-adic2}) in the space of functions supported in the ball $|x|_p\le p$ which are constant in the both potential wells and are constant at the complement to these potential wells in the ball $|x|_p\le p$. This space has the dimension either 2 (for $p=2$) or 3 (for $p>2$). The key observation is that the operators of multiplication by the functions $k_1(x)$, $k_2(x)$, $e^{\beta U_1(x)}$, $e^{\beta U_2(x)}$ map this finite dimensional space into itself. Therefore the restriction of the system (\ref{p-adic1}), (\ref{p-adic2}) to the considered space of functions takes the form of the finite system of ordinary differential equations.

For simplicity we restrict our discussion to the case $p=2$. The case $p>2$ can be discussed analogously.

An arbitrary function from the space under consideration can be put in the form
\begin{equation}\label{wavelet_expansion}
C_0\Omega(|px|_p)+C_1\psi(x)= \Omega(|px|_p)(C_0+C_1\psi(x))=
$$
$$
=e^{-\beta U_1(x)}(A_0 + A_1\psi(x))= e^{-\beta U_2(x)}(B_0+ B_1\psi(x)).
\end{equation}
where
$$
\psi(x)=\Omega(|x|_p)-\Omega(|x-a|_p)
$$
is a wavelet.
The coefficients $A_i$, $B_i$, $C_i$ are linearly related.
One gets for the matrix elements of the operators of multiplication by $e^{\beta U_1(x)}$, $e^{\beta U_2(x)}$
$$
e^{\beta U_1(x)}\Omega(|px|_p)={1+e^{\beta U_1}\over 2}\Omega(|px|_p)-{1-e^{\beta U_1}\over 2}\psi(x),
$$
$$
e^{\beta U_1(x)}\psi(x)=-{1-e^{\beta U_1}\over 2}\Omega(|px|_p)+{1+e^{\beta U_1}\over 2}\psi(x),
$$
$$
e^{\beta U_2(x)}\Omega(|px|_p)={1+e^{\beta U_2}\over 2}\Omega(|px|_p)+{1-e^{\beta U_2}\over 2}\psi(x),
$$
$$
e^{\beta U_2(x)}\psi(x)={1-e^{\beta U_2}\over 2}\Omega(|px|_p)+{1+e^{\beta U_2}\over 2}\psi(x).
$$
Analogously for the operators of multiplication by  $k_i(x)$ we get
$$
k_1(x)\Omega(|px|_p)={k_1\over 2}\Omega(|px|_p)+{k_1\over 2}\psi(x),
$$
$$
k_1(x)\psi(x)={k_1\over 2}\Omega(|px|_p)+{k_1\over 2}\psi(x),
$$
$$
k_2(x)\Omega(|px|_p)={k_2\over 2}\Omega(|px|_p)-{k_2\over 2}\psi(x),
$$
$$
k_2(x)\psi(x)=-{k_2\over 2}\Omega(|px|_p)+{k_2\over 2}\psi(x).
$$

Application of expansions of the form (\ref{wavelet_expansion}) allows us to investigate the ultrametric reaction--diffusion in the potential since, in particular
$$
D_{x}^{\lambda\beta}e^{\beta U_{1}(x)}\cdot e^{-\beta U_1(x)}(A_0+A_1\psi(x))=A_1\psi(x).
$$
We will use the following expansions for $f_i(x,t)$
$$
f_{1}(x,t)=e^{-\beta U_1(x)}\left(A_0^{(1)}(t)+A_1^{(1)}(t)\psi(x)\right)=C_0^{(1)}(t)\Omega(|px|_p)+C_1^{(1)}(t)\psi(x),
$$
$$
f_{2}(x,t)=e^{-\beta U_2(x)}\left(B_0^{(2)}(t)+B_1^{(2)}(t)\psi(x)\right)=C_0^{(2)}(t)\Omega(|px|_p)+C_1^{(2)}(t)\psi(x).
$$

We get the following system of the four linear ordinary differential equations
for  $C_j^{(i)}(t)$:
$$
{d\over d t} \left( C_0^{(1)}(t)+C_0^{(2)}(t)\right)=0;
$$
$$
{d\over d t} \left( C_1^{(1)}(t)+C_1^{(2)}(t)\right)=
$$
$$
=-\left[-{1-e^{\beta U_1}\over 2} C_0^{(1)}(t)+{1+e^{\beta U_1}\over 2} C_1^{(1)}(t)+ {1-e^{\beta U_2}\over 2} C_0^{(2)}(t)+ {1+e^{\beta U_2}\over 2} C_1^{(2)}(t)\right];
$$
$$
{d\over d t} \left( C_0^{(1)}(t)-C_0^{(2)}(t)\right)=-\left[C_0^{(1)}(t)k_1+C_1^{(1)}(t)k_1-C_0^{(2)}(t)k_2+C_1^{(2)}(t)k_2\right];
$$
$$
{d\over d t} \left( C_1^{(1)}(t)-C_1^{(2)}(t)\right) =-\left[C_0^{(1)}(t)k_1+C_1^{(1)}(t)k_1+C_0^{(2)}(t)k_2-C_1^{(2)}(t)k_2\right]-
$$
$$
-\left[-{1-e^{\beta U_1}\over 2} C_0^{(1)}(t)+{1+e^{\beta U_1}\over 2} C_1^{(1)}(t)- {1-e^{\beta U_2}\over 2} C_0^{(2)}(t)- {1+e^{\beta U_2}\over 2} C_1^{(2)}(t)\right].
$$
This system describes the relaxation to the stationary state (\ref{f_1_stationary}), (\ref{f_2_stationary}).

\medskip

The space of locally constant functions with the support in the ball  $|x|_p\le p$ is an orthogonal sum of the space investigated above and the space of wavelets with the support in subballs in $|x|_p\le p$ or in the subset of  $|x|_p\le p$ which does not contain the potential wells. The dynamics described by equations (\ref{p-adic1}), (\ref{p-adic2}) with the initial state equal to any of the mentioned wavelets is described by exponential decay. Let us consider the following cases.

A) The initial condition for (\ref{p-adic1}), (\ref{p-adic2}) is proportional to the wavelet  $\psi_{J}$:
\begin{equation}\label{iniwave}
f_i(x,0)=f_i(0)\psi_J(x),
\end{equation}
the ball $J$ lies in the ball $|x|_p\le p$ but does not belong to the potential wells $|x|_p\le 1$, $|x-a|_p\le 1$. We get
$$
{d\over d t} f_{i}(x,t)=-D_{x}^{\lambda\beta} f_i(x,t)=-\lambda_J f_i(x,t),
$$
where $D_{x}^{\lambda\beta}\psi_{J}=\lambda_J\psi_{J}$. Therefore $f_1(x,t)$, $f_2(x,t)$ decay as $e^{-\lambda_J t}f_i(0)\psi_J(x)$.

B) Let us consider the initial condition (\ref{iniwave}) where the ball $J$ lies in the ball $|x|_p\le 1$ (one of the potential wells). We get
$$
f_{1}(x,t)=f_1(0)e^{-t\left[k_{1}+e^{\beta U_{1}}\lambda_J\right]}\psi_{J}(x);
$$
$$
f_2(x,t)=\eta f_1(x,t)+\psi_J(x)e^{-\lambda_J t}(f_2(0)-\eta f_1(0)),\quad \eta= { k_{1}+e^{\beta U_{1}}\lambda_J\over k_{1}+(1+e^{\beta U_{1}})\lambda_J}.
$$

C) The initial condition (\ref{iniwave}) with the ball $J$ lying in the second potential well $|x|_p\le 1$ $|x-a|_p\le 1$. We get the above decay of the initial condition with the transposed indices 1 and 2.

\section{Conclusion}

One of the approaches to dynamics of a molecular machine is to represent this dynamics as a cycle containing several stages (corresponding to the different binding with ligands or the different electronic states of the molecule) where each of the stages in the cycle is described by the reaction--diffusion equation with drift.

This model although giving a reasonable description of a molecular motor, has some drawbacks. First, it is able to give only numerical description of the dynamics (no analytical results are possible). Second, the protein dynamics (in particular, dynamics of molecular motors) are characterized by multiple reaction pathways which are ignored in the above model. Third, it is known that the protein dynamics is related to anomalous diffusion \cite{Frauenfelder4}, \cite{Hatom} and dynamics on complex landscapes of energy.

The alternative approach is to consider the discrete stochastic models \cite{M2} where the parallel pathways are taken into account. In these models, in order to approximate the continuous motion of a protein one has to consider a large number of intermediate states with some distribution of transition times between these states.

The approximation of basin to basin kinetics for dynamics on complex energy landscapes (in particular for protein dynamics) can be considered as a particular type of discrete stochastic models. Relevance of this approximation is well established for short peptides \cite{BeckerKarplus} and can be conjectured for larger proteins \cite{ABKO}. One can consider the limit of infinitely complex landscape of energy as some special form of continuous limit for model of basin to basin kinetics \cite{ABK} where the limit of generator of dynamics is taken not in the real but in ultrametric space (i.e. in the limit instead of real standard or anomalous diffusion we obtain an ultrametric diffusion). Mathematically this is justified by $p$-adic parametrization of the Parisi matrix (and more general models of similar kind).

Using this idea in the present paper we propose the model which combines the properties of the both continuous and discrete models by application of ultrametric ($p$-adic) reaction--diffusion with a drift, where the diffusion is generated by an ultrametric pseudodifferential operator. One of the advantages of this model is that it is exactly solvable, i.e. we give the analytical expression for the stationary state of the molecular machine (where the machine performs transitions in the stationary regime) and for the relaxation of the machine to the stationary state.

This models unifies the properties of continuous and discrete models of molecular motors in the following sense.
The $p$-adic conformational coordinate  describes the hierarchy of local minima of the energy landscape of a protein. The distribution of transition times for conformational rearrangements is described by the function of $p$-adic argument (the kernel of the diffusion generator). Since ultrametric diffusion is a jump process our model is discrete, but we take into account all possible transitions to arbitrarily small distances (in this sense the model is continuous).

\bigskip\bigskip

\noindent{\bf Acknowledgments}\qquad
The paper was partially supported by the grant ''Mathematical
Modeling of Complex System'' of the Faculty of Natural Science and Technology
of Linnaeus University.
One of the authors (S.K.) gratefully
acknowledges being partially supported
by the Program of the Department of Mathematics of the Russian
Academy of Sciences ''Modern problems of theoretical mathematics''.

\end{document}